\documentclass[aps,prd,reprint,preprintnumbers,groupedaddress]{revtex4-1}
\usepackage[latin1]{inputenc}
\usepackage{amsmath}
\usepackage{amsfonts}
\usepackage{amssymb}
\usepackage{slashed}
\usepackage{color}
\usepackage{soul}
\usepackage{graphicx}
\begin{document}
\title{Exact Solutions to the Fermion Propagator Schwinger-Dyson Equation\\ in Minkowski space with on-shell Renormalization for Quenched QED}
\author{Shaoyang Jia }
\email{sjia@email.wm.edu}
\affiliation{Physics Department, College of William \& Mary, Williamsburg, VA 23187, USA}
\author{ M.R. Pennington }
\email{michaelp@jlab.org}
\affiliation{Theory Center, Thomas Jefferson National Accelerator Facility, Newport News, VA 23606, USA\\Physics Department, College of William \& Mary, Williamsburg, VA 23187, USA ~and\\School of Physics \& Astronomy, Glasgow University, Glasgow G12 8SU, UK}

\begin{abstract}
With the introduction of a spectral representation, the Schwinger-Dyson equation (SDE) for the fermion propagator is formulated in Minkowski space in QED. After imposing the on-shell renormalization conditions, analytic solutions for the fermion propagator spectral functions are obtained in four dimensions with a renormalizable version of the Gauge Technique anzatz for the fermion-photon vertex in the quenched approximation in the Landau gauge. Despite the limitations of this model, having an explicit solution provides a guiding example of the fermion propagator with the correct analytic structure. The Pad\'{e} approximation for the spectral functions is also investigated. 
\end{abstract}
\maketitle
\section{Introduction}
The Bethe-Salpeter/Faddeev approach to hadronic physics requires quark and gluon propagators as input conditions \cite{Bashir:2012fs,Cloet:2013jya,Eichmann:2016yit}. These propagators can, in principle, be solved from their Schwinger-Dyson equations (SDEs) \cite{Schwinger:1951ex,Schwinger:1951hq,Dyson:1949ha}. The gluon self-coupling inevitably complicates the QCD equations \cite{Strauss:2012dg}. Therefore we take a more tractable approach by studying general structures and solutions in strongly coupled QED, particularly for the fermion propagator. 

The SDEs for the fermion propagator involve the photon propagator and the fermion-photon vertex. Since through its own SDE, the fermion-photon three-point function couples to higher n-point functions, solving the propagator equations requires an ansatz for the vertex in order to truncate the infinitely coupled system. The rainbow-ladder \cite{Bhagwat:2003vw} truncation, although simple and intuitive, fails to respect the Ward-Green-Takahashi identity (WGTI). The Ball-Chiu vertex~\cite{Ball:1980ay}, with the correct longitudinal components of the three-point function, respects the WGTI, but violates multiplicative renormalizability. Adding transverse pieces to the Ball-Chiu vertex recovers multiplicative renormalizablity \cite{Curtis:1990zs,Kizilersu:2009kg}, but does not ensure solutions for the propagator SDEs are gauge covariant \cite{Kizilersu:2013hea}.

Importantly, the WGTI, multiplicative renormalizability and gauge covariance provide principles for a consistent truncation of the SDEs for propagators of gauge theories. To these we add the analytic structure of the fermion propagator which constrains the singularities of the fermion-photon vertex. It is therefore highly desirable to obtain one set of illustrative solutions to the propagator SDE in four dimensions with the correct analytic structure. We do this here in a modeling with a purely \lq\lq longitudinal'' vertex \cite{Delbourgo:1977jc,PhysRev.130.1287,PhysRev.135.B1398,PhysRev.135.B1428} in quenched QED. With this modeling, solutions can be found explicitly in analytic form. 

This article is organized as follows. Sect.~\ref{ss:spec_rep} introduces the spectral representation for the fermion propagator with its two scalar spectral functions. Sect.~\ref{SS:div_analysis} discusses the requirement for removing loop divergences in the fermion propagator SDEs by renormalization conditions. Sect.~\ref{ss:solution_xi0} applies the on-shell renormalization conditions, and solves for the fermion propagator spectral functions in the Landau gauge. Sect.~\ref{ss:Pade} provides an approximation for the solutions obtained in Sect.~\ref{ss:solution_xi0}. In Sect.~\ref{ss:summary} we summarize.
\section{Spectral representation for propagators\label{ss:spec_rep}}
\subsection{Scalars}
\begin{figure}
\centering
\includegraphics[width=0.8\linewidth]{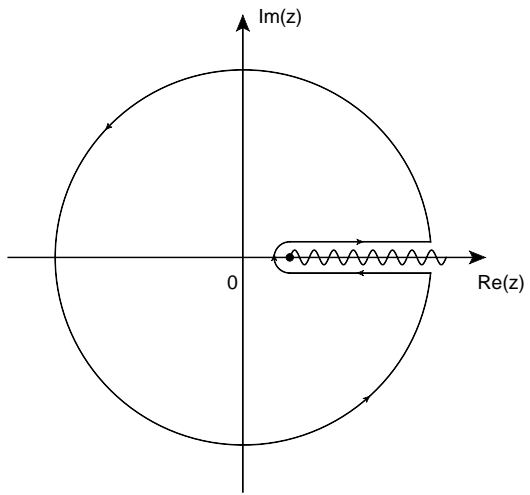}
\caption{Illustration of the analytic structure of the propagator function with dimensionless variables in the complex plane. The contour corresponds to evaluating Eq.~\eqref{eq:KLSR_scalar} using the Cauchy integration theorem.}
\label{fig:analytic_fz}
\end{figure}

A knowledge of analytic properties of propagators is required in order to solve the Schwinger-Dyson equations (SDEs) for the propagators in Minkowski space. Let us begin with the K\"{a}ll\'{e}n-Lehmann spectral representation for scalar particle propagators $D(p^2)$.  $D(p^2)$ is real for spacelike momentum $p^2<0$. When $p^2>0$, $D(p^2)$ becomes complex due to the production of real particles through quantum loop corrections.  The spectral representation is obtained by noting that the dressed propagator $D(p^2)$ can be written as a linear combination of free-particle propagators with different mass, so that
\begin{equation}
D(p^2)\,=\,\int_{m^2}^{\infty}\,ds\;\dfrac{\rho(s)}{p^2-s+i\epsilon}\, ,\label{eq:KLSR_scalar}
\end{equation}
where $\rho(s)$ is the spectral function of $D(p^2)$. The $i\epsilon$ is essential as the propagator is expected to develop a branch cut for $p^2>m^2$. For the bare spectral function, canonical quantization requires~\cite{Weinberg:1995mt} 
\begin{equation}
\int_{m^2}^{+\infty}\,ds\,\rho(s)=1 \; .
\end{equation}
Assuming the renormalization for this scalar propagator relates the bare to renormalized quantity by $D_B(p^2)=Z\,D_R(p^2)$, one can easily derive that for renormalized spectral function $\int_{m^2}^{+\infty}\,ds\,\rho_R(s)=Z^{-1}$.
For Eq.~\eqref{eq:KLSR_scalar} to converge, $\rho(s)$ must go to zero as $|s| \to \infty$. If this integral does not converge, one can make subtractions by writing for instance
\begin{align}
D(p^2)\,& =\,D(p_0^2)\,  -\int_{m^2}^{\infty} ds \dfrac{(p^2-p_0^2)\,\rho(s)}{(p^2-s+i\epsilon) (p_0^2-s+i\epsilon)}\, .\label{eq:subtraction_scalar}
\end{align}
Though in the cases we study here, renormalization ensures the integrals converge and no subtraction is needed. One can safely assume that the propagator function $D(p^2)$ is holomorphic everywhere for complex $p^2$ except for the branch cut and perhaps a finite number of poles on the positive real axis, as illustrated in Fig.~\ref{fig:analytic_fz}.

To understand such structures, consider perturbative calculations, which have been performed to all orders. Since Feynman rules apply for each diagram encountered, we can use standard techniques to combine all denominators. Then after evaluating the loop integrals, the resulting functions are expressed as integrations over the Feynman parameters. The propagator is singular when the combined denominator vanishes. This corresponds to $p^2>p^2_{\mathrm{th}}$, where $p^2_{\mathrm{th}}$ is the particle production threshold. The numerators do not modify the singularities because they are only polynomials of $p^2$. The propagator function $D(p^2)$ and its spectral function $\rho(s)$ are naturally interconnected. The spectral function $\rho(s)$ is given by a sum of delta functions representing single particle poles plus the discontinuity of the propagator function $D(p^2)$ across its right hand cut:
\begin{align}
\rho(s)& =-\dfrac{1}{2\pi i}\, [D(s+i\epsilon) - D(s-i\epsilon)]\nonumber\\
& =\,-\dfrac{1}{\pi}\mathrm{Im}\{D(s+i\epsilon) \}.
\end{align}
The advantage of the spectral representation is that it determines the propagator function everywhere in the complex momentum plane, up to a possible subtraction constant $D(p_0^2)$, as in Eq.~\eqref{eq:subtraction_scalar}.

\subsection{Fermions}
The Dirac structure of a fermions means its propagator involves both the unit and $\gamma$ matrices. The algebra defined by $\{\gamma^\mu,\gamma^\nu\} = 2 g^{\mu\nu}$ with $g^{\mu\nu}$ the metric tensor, is usually represented by  $n\times n$ matrices (with $n=4$) in 3 and 4-dimensions. The fermion propagator carrying momentum $p$, denoted by $S_F(p)$, is generally written in terms of two scalar functions, the wavefunction renormalization ${\cal F}(p^2)$ and the mass function ${\cal M}(p^2)$, or equivalently in terms of functions $A(p^2)$ and $B(p^2)$, such that
\begin{eqnarray}\label{eq:dressingdefined} 
S_F(p)&=&\dfrac{{\cal F}(p^2)}{\slashed{p} - {\cal M}(p^2)}\;,\\[1mm]
S_F^{\,-1}(p^2)&=&A(p^2)\,\slashed{p} + B(p^2)\; .
\end{eqnarray}
The bare propagator is the special case with ${\cal F} =1$ and ${\cal M} = m$. To replicate this Dirac structure requires two spectral functions, $\rho_j(s)$ with $j=1,2$, so that:
\begin{eqnarray}
&&S_F(p)\,=\,\slashed{p}S_1(p^2)+S_2(p^2)\nonumber\\
&&=\slashed{p}\,\int_{m^2}^{\infty} ds \dfrac{\rho_1(s)}{p^2-s+i\epsilon} + \int_{m^2}^{\infty} ds \dfrac{\rho_2(s)}{p^2-s+i\epsilon}.
\end{eqnarray}
Following Ref.~\cite{Delbourgo:1977jc}, we take the square root of the integration variable such that $W=\sqrt{s}$. Information carried by $\rho_1(s)$ and $\rho_2(s)$ can then be combined into one function, which is particularly convenient for calculating loop integrals,
\begin{equation}
\rho(W)=\mathrm{sign}(W)\,[W\rho_1(W^2)+\rho_2(W^2)].
\end{equation}
Then, the  spectral representation of the fermion propagator can be written as
\begin{equation}
S_F(p)\,=\,\int_{|W|\geq m}^{+\infty}dW\dfrac{\rho(W)}{\slashed{p}-W+i\epsilon~\mathrm{sign}(W)}.
\end{equation}
Because the quenched approximation will be used throughout this article, the issue of the spectral representation for the photon propagator need not be discussed.

The spectral function $\rho(W)$ is renormalization scheme dependent. Using the Gauge Technique in its original form~\cite{Delbourgo:1977jc} (to be discussed in more detail in the next Section), and a renormalization scheme corresponding to
\begin{equation}
1\,=\,Z_2\int dW\rho(W),\quad mZ_m\,=\,Z_2\int dW~W\rho(W),\label{eq:Z2_Zm_Delbourgo} 
\end{equation}
the fermion propagator spectral function $\rho(W)\,={}\,{\delta(W-m)+r(W)}$ has been solved. The result is given by Eq.~(20) of Ref.~\cite{Delbourgo:1977jc} as
\begin{align}
r(W)&=\,-\mathrm{sign}(W)\theta(W^2-m^2)\dfrac{2a}{W}\left(\dfrac{W^2-m^2}{\mu^2}\right)^{-2a}\nonumber\\
&\quad\times\dfrac{m^2}{W^2-m^2}\Bigg\{~_2F_1\left(-a,-a;-2a;1-\dfrac{W^2}{m^2}\right)\nonumber\\
&\hspace{0.75cm} +\dfrac{W}{m}~_2F_1\left(-a, 1-a, -2a, 1-\dfrac{W^2}{m^2}\right)\Bigg\},\label{eq:rW_Delbourgo}
\end{align}	
where $a\,=\,3\alpha/(4\pi)$. Eq.~\eqref{eq:rW_Delbourgo} results in momentum space functions more singular than the free-particle propagators. This can be verified by applying Eq.~(15.3.6) of Ref.~\cite{abramowitz1964handbook} to Eq.~(21) of Ref.~\cite{Delbourgo:1977jc} in the $p^2\rightarrow m^2$ limit. In the next two sections, using an on-shell renormalization scheme with a modification to the Gauge Technique required by renormalizability, a different solution for $r(W)$ is obtained.
\section{Renormalization of fermion propagator SDE loop infinities\label{SS:div_analysis}}
A crucial relation between the fermion propagator and the fermion-boson vertex is imposed by gauge invariance in the form of the Ward-Green-Takahashi identity. This requires that $\Gamma^\mu(k,p)$, the grey vertex in Fig.~2, satisfies
\begin{equation}
q^\mu\,\Gamma_\mu(k,p)\,=\,S_F^{\,-1}(k) - S_F^{\,-1}(p)\; ,
\end{equation}
with $q = k-p$. Importantly, this has a non-singular limit when $q \to 0$, viz. the Ward identity, so that $\Gamma^\mu(p,p) = \partial S_F^{\,-1}(p)/\partial p_\mu$.  As is well-known these constraints are satisfied by the Ball-Chiu vertex, $\Gamma_{BC}^\mu(k,p)$,~\cite{Ball:1980ay}
where
\begin{align}
\Gamma^\mu_{BC}(k,p)=&\frac{1}{2}\,\left( \frac{1}{{\cal F}(k^2)} + \frac{1}{{\cal F}(p^2)}\right) \gamma^\mu \nonumber\\
&+\, \frac{1}{2}\,\left( \frac{1}{{\cal F}(k^2)} \,-\, \frac{1}{{\cal F}(p^2)}\right) \frac{(\slashed{k}+\slashed{p}) (k+p)^\mu}{k^2-p^2}\nonumber\\[1mm]
&-\, \left( \frac{{\cal M}(k^2)}{{\cal F}(k^2)} + \frac{{\cal M}(p^2)}{{\cal F}(p^2)}\right) \frac{(k+p)^\mu}{k^2-p^2}.
\end{align}
To this any transverse vertex $\Gamma^\mu_T(k,p)$ can be added, provided it satisfies $q_\mu \Gamma^\mu_T(k,p) = 0$ and $\Gamma_T^\mu(p,p) = 0$.

One specific spectral construction of the vertex satisfying the longitudinal Ward-Green-Takahashi identity is provided by the Gauge Technique~\cite{Delbourgo:1977jc}. This ansatz is naturally linear in the spectral function $\rho(s)$:
\begin{equation}
S_F(p)\Gamma^\mu(k,p)S_F(p)\,=\int dW\, \dfrac{1}{\slashed{k}-W} \gamma^\mu \dfrac{1}{\slashed{p}-W}\, \rho(W).\label{eq:GT}
\end{equation}

Aside from the WGTI, remormalizability also constrains the vertex. While multiplicative renormalizability conditions are rather strong, the ability to remove divergences from the loop integral in fermion propagator SDE, Fig.~2, also constrains the fermion-photon vertex in a weaker sense. To differentiate this from multiplicative renormalizability, we call this loop-renormalizability. 

\begin{figure*}
\centering
\includegraphics[width=0.7\linewidth]{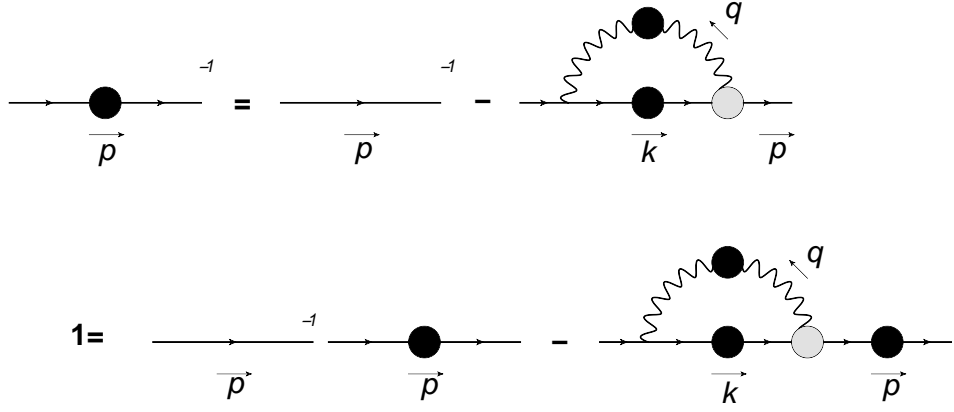}
\caption{Diagrammatic representation of fermion propagator SDE in propagator form (upper) and spectral form (lower).}
\label{fig:DSE_text}
\end{figure*}
The fermion propagator satisfies the SDE displayed in the upper half of Fig.~2. This requires that
\begin{equation}\label{eq:SDE} 
S_F^{\,-1}(p)\,=\, {S^0}_F^{-1}(p)\,
+\, i e^2\,\int\, d{\underline k}\,\gamma^\nu S_F(k) \Gamma^\mu(k,p) D_{\mu\nu}(q)\, ,
\end{equation}
where $d{\underline k} = d^dk/(2\pi)^d$ in $d$-dimensions, and $D^{\mu\nu}$ is the photon propagator carrying momentum $q=k-p$ as in Fig.~2. In the quenched approximation, we have 
\begin{equation}
D_{\mu\nu}(q)\,=\,\dfrac{1}{q^2+i\epsilon}\left[g_{\mu\nu}+(\xi-1)\dfrac{q_\mu q_\nu}{q^2}\right]\; ,
\end{equation}
in a covariant gauge specified by the parameter $\xi$.
Multiplying Eq.~\eqref{eq:SDE} throughout by the full fermion-propagator, we have
\begin{align}
1\,&=\,(\slashed{p} - m) S_F(p)\,\nonumber\\
& \quad+\, i e^2 \,\int\, d{\underline k}\,\gamma^\nu S_F(k) \Gamma^\mu(k,p) S_F(p) D_{\mu\nu}(q)\, .\label{eq:SDE2}
\end{align}
A spectral representation such as Eq.~\eqref{eq:GT} allows the loop integral in Eq.~\eqref{eq:SDE2} to be evaluated exactly. Substituting the ansatz Eq.~\eqref{eq:GT} of the Gauge Technique into Eq.~\eqref{eq:SDE2} yields
\begin{align}
1\,& =\,(\slashed{p}-m)S_{F}(p)\,+\,ie^2\int\, d\underline{k}\,\int dW\,\gamma^\nu\dfrac{1}{\slashed{k}-W}\nonumber\\
& \quad\hspace{2.75cm} \times\,\gamma^\mu\,\dfrac{1}{\slashed{p}-W}\,D_{\mu\nu}(q)\,\rho(W)\,.\label{eq:SDE_fermion_bare}
\end{align}

We define the following functions linear in the spectral function $\rho(W)$,
\begin{align}
&\quad \sigma_1(p^2)+\slashed{p}\sigma_2(p^2)\,\nonumber\\
&=\,ie^2\,\int\, d\underline{k}\int  dW \gamma^\nu\dfrac{1}{\slashed{k}-W}\gamma^\mu \dfrac{1}{\slashed{p}-W} D_{\mu\nu}(q) \rho(W).\label{eq:GT_unmod}
\end{align}
Adding transverse pieces to Eq.~\eqref{eq:GT} has the potential to modify the divergences in Eq.~\eqref{eq:GT_unmod}.
After renormalization, Eq.~\eqref{eq:SDE_fermion_bare} becomes
\begin{subequations}\label{eq:SDE_R}
\begin{align}
& Z_2^{-1}+mZ_mS_2(p^2)\,=\,p^2S_1(p^2)+\sigma_1(p^2) \label{eq:SDE_R_Z2}\\[1mm] 
& mZ_mS_1(p^2)\,=\,S_2(p^2)+\sigma_2(p^2).\label{eq:SDE_R_Zm}
\end{align}
\end{subequations}
Eq.~\eqref{eq:SDE_R} couples the fermion propagator functions $S_{j}(p^2)$. In order to derive the corresponding equations for the spectral functions $\rho_{j}(s)$, we need to find out how to generate these $p^2$ dependences in Eq.~\eqref{eq:SDE_R} from the free-particle propagator by taking imaginary parts. However, as will be demonstrated in the following section, we are faced with a more immediate problem that the Gauge Technique ansatz for the fermion-photon vertex yields $\sigma_{j}(p^2)$, with the divergent parts that cannot be removed by renormalization conditions. 

It is a fundamental principle that QED is renormalizable. This serves as an important criterion for truncating the fermion SDE~\cite{Curtis:1990zs}. As already remarked, the Ball-Chiu vertex~\cite{Ball:1980ay}, being only longitudinal, although satisfying the Ward-Green-Takahashi identity for fermion-photon vertex, fails to ensure multiplicative renormalizability. While the ansatz of the Gauge Technique does not fulfill this requirement either, we can make the solutions satisfy the weaker condition of loop-renormalizability, which we now introduce.

The principle of loop-renormalization is best illustrated by considering the SDE for propagator functions. We define the fermion self-energy as 
\begin{eqnarray}
\Sigma_1(p^2)\slashed{p}+\Sigma_2(p^2)&=&(\sigma_1+\slashed{p}\sigma_2)S_F^{-1}\nonumber\\
&&\hspace{-2.8cm}=\dfrac{\sigma_1S_1-\sigma_2S_2}{p^2S_1^2-S_2^2}\slashed{p}+\dfrac{p^2S_1\sigma_2-\sigma_1S_2}{p^2S_1^2-S_2^2}.\label{eq:def_Sigma_j_fermion_self_energy}
\end{eqnarray}
In terms of the previously introduced dressing functions in Eq.~\eqref{eq:dressingdefined}, we have ${p^2S_1^2(p^2)-S_2^2(p^2)\,=\,\mathcal{F}(p^2)S_1(p^2)}$. Next, using the Gauge Technique in the quenched approximation, we evaluate the integrals as functions of $d=4-2\epsilon$, and expand the answers in powers of $\epsilon$. Then the divergent parts of the fermion self-energy are given by
\begin{subequations}\label{eq:Sigma_j_ori}
\begin{align}
\Sigma_1(p^2)&  \,=\,-\dfrac{3\alpha}{4\pi\epsilon}\left[1-\dfrac{(1+\xi/3)}{Z_2\mathcal{F}(p^2)}\right]+\mathcal{O}(\alpha\epsilon^0),\label{eq:Sigma_1_ori}\\[1.5mm]
\Sigma_2(p^2)& =-\dfrac{3\alpha}{4\pi\epsilon}\,\dfrac{(1+\xi/3){\cal M}(p^2)}{Z_2\mathcal{F}(p^2)}+\mathcal{O}(\alpha\epsilon^0).\label{eq:Sigma_2_ori}
\end{align}
\end{subequations}
Recall the renormalized SDE for the fermion propagator in its original form are given by
\begin{subequations}\label{eq:SDE_F_MF}
\begin{align}
& \dfrac{1}{Z_2\mathcal{F}(p^2)}\,=\,1+\Sigma_1(p^2),\label{eq:SDE_F}\\[1.5mm]
& \dfrac{{\cal M}(p^2)}{Z_2\mathcal{F}(p^2)}\,=\,m_RZ_m-\Sigma_2(p^2)\, .\label{eq:SDE_MF}
\end{align}
\end{subequations}
At first sight, one might renormalize Eq.~\eqref{eq:SDE_F} by multiplying by ${\cal F}(p^2)$ to give an expression for $Z_2^{\,-1}$. Since $Z_2$ is independent of momentum, it is fixed by its value at $p^2=\mu^2$, to give  
\begin{equation}
\dfrac{\mathcal{F}(\mu^2)}{\mathcal{F}(p^2)}\,=\,\dfrac{1+\Sigma_1(p^2)}{1+\Sigma_1(\mu^2)}\, .\label{eq:ratio_F}
\end{equation}
However, this equation has divergences that have not been removed. Substituting Eq.~\eqref{eq:Sigma_1_ori} into the right-hand side of Eq.~\eqref{eq:ratio_F} fails to reproduce the left-hand side in the small $\epsilon$ limit. Thus the fermion SDE with the Gauge Technique Eq.~\eqref{eq:GT_unmod} is not loop-renormalizable.

Let us imagine that instead of Eq.~\eqref{eq:Sigma_j_ori} the divergences for the fermion self-energy can be written as
\begin{subequations}\label{eq:loop_renormalizability} 
\begin{align}
\Sigma_1(p^2)& =\dfrac{\alpha\xi}{4\pi\epsilon} \dfrac{1}{Z_2\mathcal{F}(p^2)}+\overline{\Sigma}_1(p^2)+\mathcal{O}(\alpha\epsilon^1),\label{eq:Sigma_1_alt}\\[1.5mm]
\Sigma_2(p^2)& =-\dfrac{\alpha\xi}{4\pi\epsilon} \dfrac{\mathcal{M}(p^2)}{Z_2\mathcal{F}(p^2)}+\overline{\Sigma}_2(p^2)+\mathcal{O}(\alpha\epsilon^1),\label{eq:Sigma_2_alt}
\end{align}
\end{subequations}
where $\overline{\Sigma}_{j}(p^2)$ are finite, hence are $\mathcal{O}(\epsilon^0)$. Here the coefficients for the divergent terms are chosen to agree with the perturbative one-loop calculation. Notice that from Eq.~\eqref{eq:Sigma_1_alt}, the divergent part of $\Sigma_1$ is homogeneous with respect to $1/Z_2\mathcal{F}(p^2)$, \textit{i.e.} the propagator term in the renormalized SDE. Such a form, Eq.~\eqref{eq:Sigma_1_alt}, can be achieved by adding transverse terms to the Gauge Technique. Satisfying Eq.~\eqref{eq:Sigma_1_alt} will make Eq.~\eqref{eq:SDE_F} loop-renormalizable. Eq.~\eqref{eq:Sigma_1_alt} allows us to rewrite Eq.~\eqref{eq:SDE_F} as
\begin{equation}\label{eq:ren_supposed1}
\left(1-\dfrac{\alpha\xi}{4\pi\epsilon}\right)\dfrac{1}{Z_2\mathcal{F}(p^2)}\,=\,1+\overline{\Sigma}_1(p^2).
\end{equation}
Now we can use renormalization conditions to eliminate $[1-\alpha\xi/(4\pi\epsilon)]Z_2^{-1}$ and obtain
\begin{equation}
\dfrac{\mathcal{F}(\mu^2)}{\mathcal{F}(p^2)}\,=\,\dfrac{1+\overline{\Sigma}_1(p^2)}{1+\overline{\Sigma}_1(\mu^2)},
\end{equation}
which is free from any divergences.

As for the other component of the renormalized SDE, combining Eq.~\eqref{eq:SDE_F} with Eq.~\eqref{eq:SDE_MF} produces,
\begin{equation}
m_R Z_m\,=\,\mathcal{M}(p^2)+\Sigma_1(p^2)\mathcal{M}(p^2)+\Sigma_2(p^2),\label{eq:SDE_M}
\end{equation} 
Next, using Eq.~\eqref{eq:loop_renormalizability}, the loop divergence is given by
\begin{equation}
\Sigma_1(p^2)\mathcal{M}(p^2)+\Sigma_2(p^2)
=\,0+\mathcal{O}(\alpha\epsilon^0).
\end{equation}
This cancellation ensures $\mathcal{M}(p^2)$ is finite. Therefore, evaluating Eq.~\eqref{eq:SDE_M} at $p^2=\mu^2$ specifies $m_R Z_m$. Consequently we have
\begin{equation}
\mathcal{M}(p^2) = \dfrac{1+\overline{\Sigma}_1(\mu^2)}{1+\overline{\Sigma}_1(p^2)}\mathcal{M}(\mu^2)+\dfrac{\overline{\Sigma}_2(\mu^2)-\overline{\Sigma}_2(p^2)}{1+\overline{\Sigma}_1(p^2)}.
\end{equation}

Summing up the previous discussion, in order to ensure the renormalized SDE for fermion propagator, Eq.~\eqref{eq:SDE_F_MF}, being renormalizable by eliminating renormalization constants $Z_2$ and $Z_m$ at $\mu^2$, divergent parts of fermion selfenergy $\Sigma_1$ and $\Sigma_2$ have to be homogeneous with respect to the propagator contribution in the SDE and cancel each other after decoupling $\mathcal{M}(p^2)$ from the Dirac scalar equation. Therefore Eq.~\eqref{eq:loop_renormalizability} needs to be satisfied. 
Unlike Eq.~\eqref{eq:GT_unmod}, substituting Eq.~\eqref{eq:loop_renormalizability} into the right-hand side of Eq.~\eqref{eq:ratio_F} produces the left-hand side. 

Notice that the fermion self-energy differs from the $\sigma_j(p^2)$ by the fermion propagator, as specified by Eq.~\eqref{eq:def_Sigma_j_fermion_self_energy}. 
Because the fermion propagator can be viewed as a linear transform with finite matrix elements, the renormalizability conditions Eq.~\eqref{eq:loop_renormalizability} indicate
\begin{subequations}\label{eq:loop_ren_sigma_j}
\begin{align}
\sigma_1(p^2)& =p^2\Sigma_1(p^2)S_1(p^2)+\Sigma_2(p^2)S_2(p^2)\nonumber\\[1mm]
&=\,\dfrac{\alpha\xi}{4\pi\epsilon}Z_2^{-1}+\overline{\sigma}_1(p^2),\label{eq:loop_ren_sigma_1}\\[1.5mm]
\sigma_2(p^2)& =\Sigma_1(p^2)S_2(p^2)+\Sigma_2(p^2)S_1(p^2)\nonumber\\[1mm]
&=\,\overline{\sigma}_2(p^2),\label{eq:loop_ren_sigma_2}
\end{align}
\end{subequations}
where $\overline{\sigma}_{j}(p^2)$ are the finite parts for $\sigma_{j}(p^2)$. Therefore we have translated Eq.~\eqref{eq:loop_renormalizability} into Eq.~\eqref{eq:loop_ren_sigma_j} as the loop-renormalizability requirement for $\sigma_j(p^2)$. Notice the divergence in Eq.~\eqref{eq:loop_ren_sigma_1} vanishes in the Landau gauge, as expected, but this does not indicate any ansatz in the Landau gauge satisfies Eq.~\eqref{eq:loop_ren_sigma_j}.
\section{Solutions for the fermion propagator spectral functions\label{ss:solution_xi0}}
\subsection{Renormalizable modification to the Gauge Technique}
To obtain knowledge of the fermion propagator spectral functions without complicating their SDEs, let us return to Eq.~\eqref{eq:GT_unmod} and first consider the Gauge Technique in the quenched approximation. The loop integral of the fermion propagator SDE, that gives the forms of Eq.~\eqref{eq:Sigma_j_ori} can easily be calculated explicitly by applying well-established perturbative procedures. After Feynman parameterization and dimensional regularization with ${d\,=\,4-2\epsilon}$, we obtain
\begin{subequations}
\begin{align}
\sigma_1(p^2)\,&=\,-\dfrac{3\alpha}{4\pi}\int ds\,\dfrac{sK(p^2,s)}{p^2-s}\,\rho_1(s) +\dfrac{\alpha\xi}{4\pi}\nonumber\\
&\quad \times\int ds\,\left(C_{div}+1+\ln\dfrac{\mu^2}{s-p^2}\right)\,\rho_1(s)\\
\sigma_2(p^2)\,&=\,-\dfrac{3\alpha}{4\pi}\int ds\,\dfrac{K(p^2,s)}{p^2-s}\,\rho_2(s)+\dfrac{\alpha\xi}{4\pi}\nonumber\\
& \quad \times\int ds\, \dfrac{1}{p^2}\left(-1+\dfrac{s}{p^2}\ln\dfrac{s}{s-p^2}\right)\,\rho_2(s),
\end{align}
\end{subequations}
where
\begin{equation}
K(p^2,s) \,=\,C_{div}+\dfrac{4}{3}+\ln\dfrac{\mu^2}{s-p^2}-\dfrac{s}{p^2}\ln\dfrac{s}{s-p^2},
\end{equation}
with $\mu$ giving the dimension of the coupling constant, which is eventually set to the fermion on-shell mass: $m$. Meanwhile, ${C_{div}\,=\,1/\epsilon-\gamma_E+\ln 4\pi}$. To make this satisfy the loop-renormalizability requirement specified by Eq.~\eqref{eq:loop_ren_sigma_j}, modifications are to replace ${K(p^2,s)}$ by
\begin{equation}
\overline{K}(p^2,s)\,=\,\dfrac{4}{3}+\left(1-\dfrac{s}{p^2}\right)\ln\dfrac{s}{s-p^2}.\label{eq:minimum_LR_GaugeTechnique}
\end{equation}
As discussed this can be accomplished by adding transverse vectors to Eq.~\eqref{eq:GT}. Within the $\overline{\mathrm{MS}}$ scheme, ${1/\epsilon-\gamma_E+\ln 4\pi}$ terms in $\sigma_{j}$ are removed altogether. Then, based on the definition of $\overline{\sigma}_{j}$ by Eq.~\eqref{eq:loop_ren_sigma_j}, Eq.~\eqref{eq:SDE_R} become
\begin{subequations}
\begin{align}
\left(1-\dfrac{\alpha\xi}{4\pi\epsilon}\right)Z_2^{-1}+mZ_mS_2&=\,p^2S_1+\overline{\sigma}_1\label{eq:SDE_oriR1}\\[1mm]
mZ_mS_1&=\,S_2+\overline{\sigma}_2\, ,\label{eq:SDE_oriR2}
\end{align} 
\end{subequations}
in a form like that supposed in Eqs.~(\ref{eq:ren_supposed1},~\ref{eq:SDE_M}).
As described in Sect.~\ref{SS:div_analysis}, renormalization at $p^2=\mu^2$ eliminates $mZ_m$, $Z_2^{-1}$ and ${\alpha\xi/(4\pi\epsilon)}$ altogether, so that
\begin{subequations}\label{eq:SDE_fqR_ori_j}
\begin{align}
&\quad\dfrac{S_2(p^2)+\overline{\sigma}_2(p^2)}{S_1(p^2)}\,=\,\dfrac{S_2(\mu^2)+\overline{\sigma}_2(\mu^2)}{S_1(\mu^2)}\,,\label{eq:SDE_fqR_ori_1}\\[1.5mm]
&\quad p^2S_1(p^2)+\overline{\sigma}_1(p^2)-S_2(p^2)\dfrac{S_2(\mu^2)+\overline{\sigma}_2(\mu^2)}{S_1(\mu^2)}\nonumber\\
& =\,\mu^2S_1(\mu^2)+\overline{\sigma}_1(\mu^2)-S_2(\mu^2)\dfrac{S_2(\mu^2)+\overline{\sigma}_2(\mu^2)}{S_1(\mu^2)}.\label{eq:SDE_fqR_ori_2}
\end{align}
\end{subequations}
Although the resulting Eq.~\eqref{eq:SDE_fqR_ori_j} appears nonlinear in the spectral functions $\rho_{j}(s)$, the mass shell renormalization will render it linear.
\subsection{On-shell renormalization conditions}
On-shell renormalization stipulates that propagator functions evaluated near the mass shell are dominated by their free-particle counterparts. Mathematically, we translate this statement into
\begin{subequations}\label{eq:OnShellCondj}
\begin{align}
& S_1(p^2)\,=\,\dfrac{1}{p^2-m^2}+P_1(p^2)\label{eq:OnShellCond1}\\
& S_2(p^2)\,=\,\dfrac{m}{p^2-m^2}+P_2(p^2),\label{eq:OnShellCond2}
\end{align}
\end{subequations}
where $P_j(p^2)$ have to be less singular than the free-particle propagator at least in the vicinity of $m^2$.
This requires that the spectral functions $\rho_j(s)$ cannot be more singular than $\delta$-functions when $s\rightarrow m^2$. 
Thus the on-shell renormalization condition Eq.~\eqref{eq:OnShellCondj} indicates
\begin{subequations}\label{eq:rhoj_delta_theta}
\begin{align}
& \rho_1(s)\,=\,\delta(s-m^2)+r_1(s),\label{eq:rho1_delta_theta}\\
& \rho_2(s)\,=\,m\delta(s-m^2)+r_2(s),\label{eq:rho2_delta_theta}
\end{align}
\end{subequations}
where $r_{j}(s)$ are supposed to be regular functions rather than distributions with exotic features. Dirac delta functions contribute to the ${(p^2-m^2)^{-1}}$ singular parts of $S_{j}(p^2)$ in Eq.~\eqref{eq:OnShellCondj} while $r_{j}(s)$ give rise to $P_{j}(p^2)$, which are expected to be at most $\ln(m^2-p^2)$ divergent when $p^2\rightarrow m^2$. Such regular functions allow the interchange of limits and integrations within the spectral representation.
Dirac delta functions in the spectral functions $\rho_{j}$ contribute to the free-propagator terms within $\overline{\sigma}_{j}(p^2)$ defined by Eq.~\eqref{eq:loop_ren_sigma_j} from the loop diagram in the fermion propagator SDE. Such contributions are given by 
\begin{subequations}\label{eq:def_bbSj}
\begin{align}
& \overline{\sigma}_1(p^2)\,=\,-\dfrac{\lambda_1\alpha}{4\pi}\,\dfrac{m^2}{p^2-m^2}+q_1(p^2),\label{eq:def_bbS1}\\
& \overline{\sigma}_2(p^2)\,=\,-\dfrac{\lambda_2\alpha}{4\pi}\,\dfrac{m}{p^2-m^2}+q_2(p^2),\label{eq:def_bbS2}
\end{align}
\end{subequations}
where $q_j(p^2)$ are no more singular than $(p^2-m^2)^{-1}$ at $p^2\rightarrow m^2$. Coefficients $\lambda_j$ are determined by the vertex ansatz.

The on-shell conditions also simplify the $\mu^2$-dependent terms in Eq.~\eqref{eq:SDE_fqR_ori_j}. Explicit steps can be found in Appendix \ref{ss:simplification_SDE_fermion}. After further separating the free-propagator terms, we rewrite Eq.~\eqref{eq:SDE_fqR_ori_j} into
\begin{subequations}\label{eq:SDE_rj_OnShell}
\begin{align}
&\quad  p^2P_1(p^2)+q_1(p^2)+\,(\lambda_2-\lambda_1)\dfrac{\alpha}{4\pi}\dfrac{m^2}{p^2-m^2}\nonumber\\[1mm]
& =\left(1-\dfrac{\lambda_2\alpha}{4\pi}\right)mP_2(p^2)+\lim\limits_{\mu^2\rightarrow m^2}\Big\{(\lambda_2-\lambda_1)\dfrac{\alpha}{4\pi}\dfrac{m^2}{\mu^2-m^2}\Big\}\nonumber\\[1mm]
& \quad  +\left(2-\dfrac{\lambda_2\alpha}{4\pi}\right)[m^2P_1(m^2)-mP_2(m^2)]\nonumber\\
& \quad +q_1(m^2)-mq_2(m^2)\label{eq:SDE_r1_OnShell}\\[2mm]
& P_2(p^2)+q_2(p^2) =\left(1-\dfrac{\lambda_2\alpha}{4\pi}\right)mP_1(p^2),\label{eq:SDE_r2_OnShell}
\end{align}
\end{subequations}
as the SDEs for the theta function parts of $\rho_j(s)$.
\subsection{Loop-renormalizable modification to the Gauge Technique}
To proceed further in the calculation, consider the minimal loop-renormalizable modification to the Gauge Technique in the quenched approximation introduced in Eq.~\eqref{eq:minimum_LR_GaugeTechnique}. In the case of the  $\overline{\mathrm{MS}}$ scheme:
\begin{subequations}\label{eq:sigma_bar_j}
\begin{align}
\overline{\sigma}_1(p^2)\,&=\,\dfrac{-3\alpha}{4\pi}\int_{m^2}^{+\infty}ds\dfrac{s\overline{K}(p^2,s)}{p^2-s}\rho_1(s)\nonumber\\
& \quad +\dfrac{\alpha\xi}{4\pi}\int_{m^2}^{+\infty}ds\left(1+\ln\dfrac{\mu^2}{s-p^2}\right)\rho_1(s)\label{eq:sigma_bar_1}\\[1.5mm]
\overline{\sigma}_2(p^2)\,&=\,\dfrac{-3\alpha}{4\pi}\int_{m^2}^{+\infty}ds\dfrac{\overline{K}(p^2,s)}{p^2-s}\rho_2(s)\nonumber\\
& \quad +\dfrac{\alpha\xi}{4\pi}\int_{m^2}^{+\infty}ds\dfrac{1}{p^2}\left(-1+\dfrac{s}{p^2}\ln\dfrac{s}{s-p^2}\right)\rho_2(s).\label{eq:sigma_bar_2}
\end{align}
\end{subequations}
From Eq.~\eqref{eq:minimum_LR_GaugeTechnique} one immediately sees ${\lambda_1\,=\,\lambda_2\,=\,4}$. We then separate the free-particle components in Eq.~\eqref{eq:sigma_bar_j} using Eqs.~(\ref{eq:rhoj_delta_theta},~\ref{eq:def_bbSj}). This results in
\begin{subequations}
\begin{align}
q_1(p^2)& =\overline{\sigma}_1(p^2)+\dfrac{\lambda_1\alpha}{4\pi}\dfrac{m^2}{p^2-m^2}\nonumber\\[1mm]
& =-\dfrac{3\alpha}{4\pi}\dfrac{m^2}{p^2}\ln\dfrac{m^2}{m^2-p^2}+\dfrac{\alpha\xi}{4\pi}\left(1+\ln\dfrac{m^2}{m^2-p^2}\right)\nonumber\\[1mm]
&\quad -\dfrac{3\alpha}{4\pi}\int_{m^2}^{+\infty}ds\dfrac{s\overline{K}(p^2,s)}{p^2-s}r_1(s)\nonumber\\[1mm]
& \quad +\dfrac{\alpha\xi}{4\pi}\int_{m^2}^{+\infty}ds\left(1+\ln\dfrac{m^2}{s-p^2}\right)r_1(s),
\end{align}
\begin{align}
q_2(p^2) & =\overline{\sigma}_2(p^2)+\dfrac{\lambda_2\alpha}{4\pi}\dfrac{m}{p^2-m^2}\nonumber\\[0mm]
& =-\dfrac{3\alpha}{4\pi}\dfrac{m}{p^2}\ln\dfrac{m^2}{m^2-p^2}\nonumber\\[0mm]
& \quad +\dfrac{\alpha\xi}{4\pi}\dfrac{m}{p^2}\Bigg\{-1+\dfrac{m^2}{p^2}\ln\dfrac{m^2}{m^2-p^2}\Bigg\}\nonumber\\[0mm]
&\quad -\dfrac{3\alpha}{4\pi}\int_{m^2}^{+\infty}ds\dfrac{\overline{K}(p^2,s)}{p^2-s}r_2(s)\nonumber\\[1mm]
& \quad +\dfrac{\alpha\xi}{4\pi}\int_{m^2}^{+\infty}ds\dfrac{1}{p^2}\left(-1+\dfrac{s}{p^2}\ln\dfrac{s}{s-p^2}\right)r_2(s).
\end{align}
\end{subequations}

Next, in order to deduce the equations for $r_j(s)$ of Eq.~\eqref{eq:rhoj_delta_theta}, we reproduce the $p^2$ dependences in Eq.~\eqref{eq:SDE_rj_OnShell} using a spectral representation as shown by Eq.~\eqref{eq:simp_sK} through Eq.~\eqref{eq:SDE_LG_OnShell}. This leads to the following integral equations for $r_j(s)$:
\begin{subequations}\label{eq:SDE_R_OnShell}
\begin{align}
& \quad\left(1-\dfrac{\alpha}{\pi}\right)[s^2r_1(s)-msr_2(s)]\nonumber\\[1mm]
&\quad+\dfrac{3\alpha}{4\pi}\left[m^2\theta(s-m^2)+\int_{m^2}^{s}ds's'r_1(s') \right]\nonumber\\[1mm]
&=\dfrac{(3-\xi)\alpha}{2\pi}s+\dfrac{\alpha\xi}{4\pi}\left[s\theta(s-m^2)+s\int_{m^2}^{s}ds'r_1(s') \right]\, , 
\end{align}
\begin{align}
& \quad \left(1-\dfrac{\alpha}{\pi}\right)[-msr_1(s)+sr_2(s)]\nonumber\\[1mm]
& \quad+\dfrac{3\alpha}{4\pi}\left[m\theta(s-m^2)+\int_{m^2}^{s}ds'r_2(s') \right]\nonumber\\[1mm]
& =\dfrac{\alpha\xi}{4\pi}\left[\dfrac{m^3}{s}\theta(s-m^2)+\dfrac{1}{s}\int_{m^2}^{s}ds's'r_2(s') \right]\, .
\end{align}
\end{subequations}
Eq.~\eqref{eq:SDE_R_OnShell} is the coupled SDE for fermion spectral functions $\rho_{j}(s)$ with a loop-renormalizable modification to the Gauge Technique in the quenched approximation. Because spectral variables $s$ and $s'$ are separable, these integral equations can be converted into differential equations by taking derivatives with respect to $s$. In the Landau gauge, only one derivative is required. Therefore in the next section, we discuss the solutions to Eq.~\eqref{eq:SDE_R_OnShell} in this gauge.
\subsection{Solutions in the Landau Gauge}
We now solve Eq.~\eqref{eq:SDE_R_OnShell} in the Landau gauge. For notational convenience, we define the coupling parameter $a$:
\begin{equation}
a\,=\,\dfrac{3\alpha/(4\pi)}{1-\alpha/\pi}.\label{eq:def_a_alpha}
\end{equation}
After setting $\xi\,=\,0$ and taking one derivative with respect to $s$, Eq.~\eqref{eq:SDE_R_OnShell} becomes
\begin{equation}
\left[
\begin{pmatrix}
s & -ms \\[1mm] 
-m & s
\end{pmatrix}
\dfrac{d}{ds}
+
\begin{pmatrix}
a+1 & -m \\[1mm] 
0 & a+1
\end{pmatrix}
\right]
\begin{pmatrix}
sr_1(s) \\[1mm] 
r_2(s)
\end{pmatrix} 
\,=\,
\begin{pmatrix}
2a \\[1mm] 
0
\end{pmatrix} .\label{eq:SDE_R_OnShell_LG_Matrix}
\end{equation}

As a first step to solving Eq.~\eqref{eq:SDE_R_OnShell_LG_Matrix}, we remove the matrix multiplying the derivative $d/ds$ by acting with the inverse matrix. We then introduce the decomposition 
\begin{equation}
{sr_1(s)\,=\,f_1(s)g_1(s)},\quad {r_2(s)\,=\,f_2(s)g_2(s)}, \label{eq:rj_decomposition}
\end{equation}
and  multiply by ${\mathrm{diag}\{g_1^{-1}(s),~g_2^{-1}(s) \}}$
to obtain coupled differential equations for $f_{j}(s)$ :
\begin{widetext}
\begin{equation}
\dfrac{d}{ds}
\begin{pmatrix}
f_1(s) \\[4mm] 
f_2(s)
\end{pmatrix} +
\begin{pmatrix}
\dfrac{g_1'(s)}{g_1(s)}+\dfrac{a+1}{s-m^2} & \dfrac{am}{s-m^2}\dfrac{g_2(s)}{g_1(s)} \\[3mm] 
\dfrac{(a+1)mg_1(s)}{(s-m^2)sg_2(s)} & \dfrac{g_2'(s)}{g_2(s)}+\dfrac{a+1-m/s}{s-m^2}
\end{pmatrix} 
\begin{pmatrix}
f_1(s) \\[4mm] 
f_2(s)
\end{pmatrix}\,=\,
\begin{pmatrix}
\dfrac{2a}{(s-m^2)g_1(s)} \\[3mm] 
\dfrac{2am}{s(s-m^2)g_2(s)}
\end{pmatrix}.\label{eq:SDE_ROS_LG_Matrix_Dcp}
\end{equation}
\end{widetext}
We fix the functions $g_j$ by requiring that the  diagonal elements of the matrix in Eq.~\eqref{eq:SDE_ROS_LG_Matrix_Dcp} vanish, such that
\begin{equation}
\dfrac{g_1'(s)}{g_1(s)}+\dfrac{a+1}{s-m^2}\,=\,0,\quad \dfrac{g_2'(s)}{g_2(s)}+\dfrac{a+1-m^2/s}{s-m^2}\,=\,0 \, .
\end{equation}
Their solution is
\begin{equation}
g_1(s)\,=\,\left(\dfrac{m^2}{s-m^2}\right)^{a+1},\quad g_2(s)\,=\,\left(\dfrac{m^2}{s-m^2}\right)^a\dfrac{m}{s},\label{eq:g_j}
\end{equation}
where integration constants are chosen so that the $f_{j}(s)$ are dimensionless. 
Eq.~\eqref{eq:SDE_ROS_LG_Matrix_Dcp} then becomes
\begin{subequations}
\begin{align}
\dfrac{d}{ds}f_1(s)& +\dfrac{a}{s}f_2(s)\,=\,\dfrac{2a}{m^2}\left(\dfrac{s-m^2}{m^2}\right)^a\label{eq:df1_f2}\\
\dfrac{d}{ds}f_2(s)&+\dfrac{(a+1)m^2}{(s-m^2)^2}\,f_1(s)\,=\,\dfrac{2a}{s-m^2}\left(\dfrac{s-m^2}{m^2}\right)^a\label{eq:df2_f1}.
\end{align}
\end{subequations}
Next, taking another derivative with respect to $s$ yields
\begin{subequations}\label{eq:ODE_fj}
\begin{align}
\dfrac{d}{ds}& s\dfrac{d}{ds}f_1(s)-\dfrac{a(a+1)m^2}{(s-m^2)^2}f_1(s)=\dfrac{2a(a+1)}{m^2}\nonumber\\
&\hspace{4.75cm}\times\left(\dfrac{s-m^2}{m^2}\right)^a\label{eq:ODE_f1}\\
\dfrac{d}{ds}& \,(s-m^2)^2\,\dfrac{d}{ds}f_2(s)-\dfrac{a(a+1)m^2}{s}\,f_2(s)\,=\,0.\label{eq:ODE_f2}
\end{align}
\end{subequations}
Eq.~\eqref{eq:SDE_R_OnShell_LG_Matrix} are now decoupled. 

The homogeneous part of Eq.~\eqref{eq:ODE_f1} can be solved by the textbook Frobenius method described in Appendix \ref{ss:Frobenius}. Adding an inhomogeneous term results in Eq.~\eqref{eq:f1_exact} that solves Eq.~\eqref{eq:ODE_f1}. There is no need to solve Eq.~\eqref{eq:ODE_f2} separately because substituting Eq.~\eqref{eq:f1_exact} into Eq.~\eqref{eq:df1_f2} gives the solution to Eq.~\eqref{eq:ODE_f2} directly.

Then, with the nontrivial expression for $c_0$ given by Eq.~\eqref{eq:c0_nontrivial}, combining Eqs.~(\ref{eq:rj_decomposition},~\ref{eq:g_j},~\ref{eq:f1_exact},~\ref{eq:f2_exact}) we obtain the solution for Eq.~\eqref{eq:SDE_R_OnShell} in the Landau gauge
\begin{subequations}\label{eq:r12_2F1}
\begin{align}
r_1(s)\,&=\,\dfrac{2a}{(a+1)s}\Bigg\{1+\dfrac{a^2}{(2a+1)}\nonumber\\
&\quad\times~_2F_1\left(a+1,a+1;2a+2;-\dfrac{s-m^2}{m^2}\right)\Bigg\} \\
\;r_2(s)\,&=\,-\dfrac{2a^2}{(2a+1)m}\nonumber\\
&\quad\times~_2F_1\left(a+1,a+2;2a+2;-\dfrac{s-m^2}{m^2}\right)
\end{align}
\end{subequations}
as the theta function parts of the fermion propagator spectral functions $\rho_{j}(s)$ in Eq.~\eqref{eq:rhoj_delta_theta}. This is our main result. These components of the spectral density are plotted as the crosses in the plots on the left hand side of Fig.~\ref{fig:alpha3n4_Pade} for $\alpha=3$ ( $a=15.9$ ) as an illustration.
\section{Pad\'{e} Approximation\label{ss:Pade}}
\begin{figure*}
\centering
\includegraphics[width=1\linewidth]{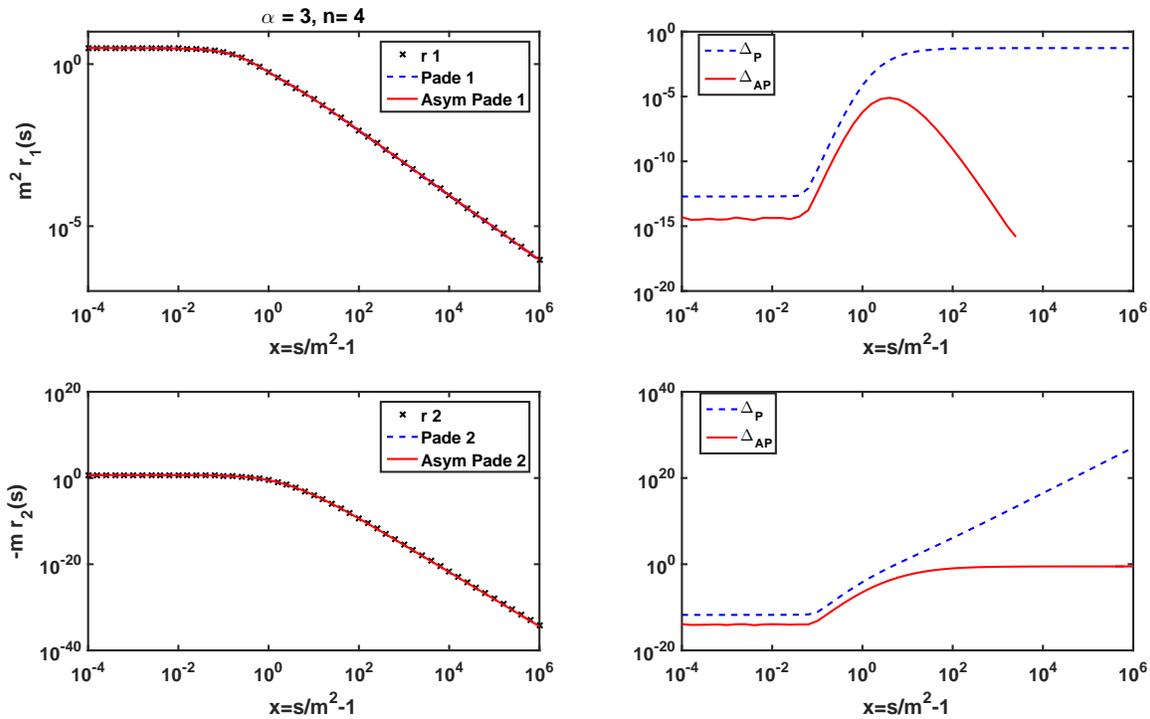}
\caption{Pad\'{e} approximations to $r_j(s)$ in Eq.~\eqref{eq:r12_2F1}. On the left the blue crosses represent the exact functions. The blue dashed lines are Pad\'{e} approximations directly applied to functions in Eq.~\eqref{eq:r12_2F1}. The red solid lines are approximations using Eq.~\eqref{eq:asym_rj}. On the right are shown the relative errors defined by $\Delta\,=\,|[r(s)-p(s)]/r(s)|$, where $r(s)$ is the exact function while $p(s)$ is the approximation.}
\label{fig:alpha3n4_Pade}
\end{figure*}

The nontrivial parts of the fermion propagator spectral functions given by Eq.~\eqref{eq:r12_2F1} are hypergeometric functions. Representing these nontrivial factors with elementary functions simplifies calculations where the fermion propagator spectral functions are used as input conditions. Since for positive $a$, the hypergeometric functions in Eq.~\eqref{eq:r12_2F1} are monotonic, one natural approximation to such functions is the Pad\'{e} approximation with the denominator polynomial being at least one degree higher than the numerator polynomial. Since the Maclaurin series for a hypergeometric function is well defined, the natural variable for the Pad\'{e} polynomials is $x\,=\,s/m^2-1$. However, the asymptotic behavior of the hypergeometric functions in Eq.~\eqref{eq:r12_2F1} cannot be reproduced by the direct application of the Pad\'{e} approximation with the variable $x$, as shown in Fig.~\ref{fig:alpha3n4_Pade}.

The asymptotic behavior of a hypergeometric function can be calculated using linear transformation formulae given by Eqs.~(15.3.3) to (15.3.9) of Ref.~\cite{abramowitz1964handbook}. However, for hypergeometric functions in $r_1(s)$ and $r_2(s)$ of Eq.~\eqref{eq:r12_2F1}, integer differences in the first two parameters require Eq.~(15.3.13) and Eq.~(15.3.14) to be applied respectively. Thus asymptotically  we have the following limiting behaviors:
\begin{subequations}\label{eq:asymp_2F1_rj}
\begin{align}
&\quad \lim\limits_{s\rightarrow +\infty}~_2F_1\left(a+1,a+1;2a+2;1-\dfrac{s}{m^2}\right)\nonumber\\
&=\dfrac{\Gamma(2a+2)}{[\Gamma(a+1)^2]}\left(\dfrac{s}{m^2}\right)^{-a-1}\left[\ln\left(\dfrac{s}{m^2} \right)+2\gamma_E-2\psi(a+1) \right]\label{eq:asymp_2F1_r1}\\
& \quad\lim\limits_{s\rightarrow+\infty}~_2F_1\left(a+1,a+2;2a+2;1-\dfrac{s}{m^2}\right)\,\nonumber\\
&=\dfrac{\Gamma(2a+2)}{\Gamma(a+1)\Gamma(a+2)}\left(\dfrac{s}{m^2}\right)^{-a-1}.\label{eq:asymp_2F1_r2}
\end{align}
\end{subequations}
Then directly from Eq.~\eqref{eq:asymp_2F1_r2}, the asymptotic behavior of $r_2(s)$ in Eq.~\eqref{eq:r12_2F1} is given by $s^{-a-1}$. Because of the presence of the logarithm, the behavior given by Eq.~\eqref{eq:asymp_2F1_r1} is only a little weaker than $s^{-a}$. Therefore the asymptotic behavior of the second term of $r_1(s)$ in Eq.~\eqref{eq:r12_2F1} can be approximated by $s^{-a-1}$. 

The analysis above for the asymptotic behaviors suggests the following approximations for $r_{j}(s)$
\begin{subequations}\label{eq:asym_rj}
\begin{align}
& r_1(s)\simeq \dfrac{2a}{(a+1)s}+\dfrac{1}{m^2}\left(\dfrac{s}{m^2}\right)^{-a}\dfrac{N_1(x)}{Q_1(x)},\label{eq:asym_r1}\\
& r_2(s)\simeq \dfrac{1}{m}\left(\dfrac{s}{m^2}\right)^{-a}\dfrac{N_2(x)}{Q_2(x)},\label{eq:asym_r2}
\end{align}
\end{subequations}
where $N_j(x)$ are polynomials of degree $n-1$, while $Q_j(x)$ are polynomials of degree $n$. Eqs.~(\ref{eq:asym_r1},~\ref{eq:asym_r2}) then allow the asymptotic behaviors of $r_j(s)$ in Eq.~\eqref{eq:r12_2F1} to be well approximated using the following Pad\'{e} approximations. 
Based on Eqs.~(\ref{eq:r12_2F1},~\ref{eq:asym_rj}), we have 
\begin{subequations}\label{eq:Pade_rj_2F1}
\begin{align}
\dfrac{N_1(x)}{Q_1(x)}&\simeq \dfrac{2a^3}{(2a+1)(a+1)}(x+1)^{a-1}\nonumber\\
& \quad\times~_2F_1\left(a+1,a+1;2a+2;-x\right),\label{eq:Pade_r1_2F1} \\
\dfrac{N_2(x)}{Q_2(x)}&\simeq -\dfrac{2a^2}{2a+1}(x+1)^{a}\nonumber\\
& \quad\times~_2F_1\left(a+1,a+2;2a+2;-x\right),\label{eq:Pade_r2_2F1}
\end{align}
\end{subequations}
where the semi-equal represents the Pad\'{e} approximations. Since the Maclaurin series for the right-hand sides of Eq.~\eqref{eq:Pade_rj_2F1} are well defined, the standard procedure of Pad\'{e} approximations applies. One example with the coupling $\alpha\,=\,3$ $(a=15.9)$ and quartic $Q_j(x)$ is displayed in Fig.~\ref{fig:alpha3n4_Pade}. 
Here the results from directly applying the Pad\'{e} approximations without the factor of $(s/m^2)^{-a}$ in Eq.~\eqref{eq:Pade_rj_2F1} are also presented. One sees on the left of the figure that with or without taking the asymptotic behavior of the hypergeometric functions into consideration, both approximations appear very good. On the right are shown the errors relative to the exact solution. We see that the asymptotic behaviors are represented much better with Eq.~\eqref{eq:asymp_2F1_rj} compared to the direct application of the Pad\'{e} approximations.

\section{Summary\label{ss:summary}}
Using a spectral representation for the fermion propagator, we have solved its Schwinger-Dyson equation with a form for the fermion-boson vertex given by a modification of the Gauge Technique of Salam, Strathdee and Delbourgo~\cite{Delbourgo:1977jc,PhysRev.130.1287,PhysRev.135.B1398,PhysRev.135.B1428}. This modification ensures the coupled fermion equations are loop-renormalizable in four dimensions. We then find an analytic solution in the Landau gauge valid in the whole complex momentum plane for the resulting equations. This solution involves hypergeometric functions $_2F_1$. The two fermion spectral functions are shown to contain  delta functions required for on-shell renormalizations plus explicit theta function terms resulting from real fermion-photon production. 

Fermion propagators are, particularly in QCD, a key building block for studies of bound states, their properties and dynamics. Having representations that are numerically reliable and easily transferable to a range of calculations are essential. Here in quenched QED we have an exact solution, and we can use this to test simpler numerical representations. We also show how once the asymptotics is extracted, the spectral function can be accurately represented by Pad\'{e} approximants. This will be useful in wider QCD studies.

Such an analytic solution serves as a basis for judging the general form of the fermion propagator that satisfies full multiplicative renormalizability as well as gauge covariance imposed by the Landau-Khalatnikov-Fradkin transformations, hopefully facilitating solutions beyond the quenched approximation when fermion-antifermion pairs can be produced. 

With the renormalization scheme in Ref.~\cite{Delbourgo:1977jc}, the solutions in different gauges with the Gauge Technique have been obtained as Eqs.~(9,~18,~19) of Ref.~\cite{Delbourgo:1980vc}. Such dependences on $\xi$ are known to be inconsistent with LKFT \cite{Jia:2016udu,Pennington:2016vxv}. We therefore do not solve Eq.~\eqref{eq:SDE_R_OnShell} in another gauge. However, if one applies the LKFT for the fermion propagator spectral functions as Eq.~(21) of Ref.~\cite{Pennington:2016vxv}, the on-shell delta function terms of Eq.~\eqref{eq:rhoj_delta_theta} are dissolved into theta functions with a positive $\xi$. While with a negative $\xi$, these terms become more singular than the delta function. In either case, the on-shell renormalization conditions are broken. The violation of on-shell conditions is accounted for by our model not being gauge covariant. Consequently, it is unable to be renormalized on-shell in two different gauges. Ans\"{a}tze compatible with the LKFT should be devoid of such difficulties. The construction of an anzatz that produces $\rho_j(s;\xi)$ with the correct gauge dependence specified by the LKFT, by adding transverse pieces to Eq.~\eqref{eq:GT_unmod}, is a subject of ongoing study. 
\begin{acknowledgments}
This material is based upon work supported by the U.S. Department of Energy, Office of Science, Office of Nuclear Physics under contract DE-AC05-06OR23177 that funds Jefferson Lab research. The authors would like to thank Professor Keith Ellis and other members of the Institute for Particle Physics Phenomenology (IPPP) of Durham University for kind hospitality during their visit when this article was finalized.
\end{acknowledgments}
\appendix
\section{Simplifications of the fermion propagator SDE\label{ss:simplification_SDE_fermion}}
Knowing the behavior of propagator functions and loop functions $\overline{\sigma}_{j}$ at $p^2\,=\,m^2$ allows us to simplify the renormalized equation Eq.~\eqref{eq:SDE_fqR_ori_j}. Explicitly, on-shell conditions simplify the renormalization constants in the following way,
\begin{subequations}
\begin{align}
& \quad \lim\limits_{p^2\rightarrow m^2}\,\dfrac{S_2(p^2)+\overline{\sigma}_2(p^2)}{S_1(p^2)}\,=\,\left(1-\dfrac{\lambda_2\alpha}{4\pi}\right)m,\\[2mm]
& \quad \lim\limits_{p^2\rightarrow m^2}\,\Big\{p^2S_1(p^2)+\overline{\sigma}_1(p^2)-\dfrac{S_2(p^2)}{S_1(p^2)}[S_2(p^2)+\overline{\sigma}_2(p^2)] \Big\}\nonumber\\
& =\lim\limits_{p^2\rightarrow m^2}\,\Bigg\{(\lambda_2-\lambda_1)\dfrac{\alpha}{4\pi}\dfrac{m^2}{p^2-m^2}\Bigg\}+q_1(m^2)-mq_2(m^2)\nonumber\\
& \quad +1+\left(2-\dfrac{\lambda_2\alpha}{4\pi}\right) [m^2P_1(m^2)-mP_2(m^2)].
\end{align}
\end{subequations}
Equations~(\ref{eq:SDE_fqR_ori_1},~\ref{eq:SDE_fqR_ori_2}) then become
\begin{subequations}
\begin{align}
& \quad p^2S_1(p^2)+\overline{\sigma}_1(p^2)\nonumber\\
& =\left(1-\dfrac{\lambda_2\alpha}{4\pi}\right)mS_2(p^2)+\lim\limits_{p^2\rightarrow m^2}\Big\{(\lambda_2-\lambda_1)\dfrac{\alpha}{4\pi}\dfrac{m^2}{\mu^2-m^2}\Big\}\nonumber\\
&\quad +q_1(m^2)-mq_2(m^2)+1\nonumber\\
& \quad +\left(2-\dfrac{\lambda_2\alpha}{4\pi}\right)[m^2P_1(m^2)-mP_2(m^2)]\label{eq:SDE_RZ1_OnShell}\\[2mm]
& \quad S_2(p^2)+\overline{\sigma}_2(p^2) =\left(1-\dfrac{\lambda_2\alpha}{4\pi}\right)mS_1(p^2),\label{eq:SDE_RZm_OnShell}
\end{align}
\end{subequations}
with $q_j(p^2)$ defined in Eq.~\eqref{eq:def_bbSj}.

The following relations are useful in obtaining the imaginary part of $q_1(q^2)$.
\begin{align}
& \dfrac{s'K(s,s')}{s-s'}\,=\,\dfrac{s'}{s-s'}\left(C_{div}+\dfrac{4}{3}+\ln\dfrac{\nu^2}{s'}\right)+\dfrac{s'}{s}\ln\dfrac{s'}{s'-s}\label{eq:simp_sK}\\[1mm]
& -\dfrac{1}{\pi}\mathrm{Im}\Big\{\dfrac{1}{s-s'+i\epsilon}\Big\}\,=\,\delta(s-s')\\[1mm]
& -\dfrac{1}{\pi}\mathrm{Im}\Bigg\{\int ds'\left(C_{div}+1+\ln\dfrac{\nu^2}{s'}\right)\rho_1(s')\Bigg\}\,=\,0\, .
\end{align}
Let us define $z\,=\,s/s'$. The logarithmic terms of $\overline{\sigma}_1$ are to be reparameterized by introducing an intermediate spectral function $\kappa(\zeta)$ for the nontrivial imaginary parts of kernel functions:
\begin{align}
& \dfrac{s'}{s}\ln\dfrac{s'}{s'-s}\,=\,\dfrac{1}{z}\ln\dfrac{1}{1-z}\,=\,-\int\,\dfrac{d\zeta}{\zeta}\, \dfrac{\theta(\zeta-1)}{z-\zeta+i\epsilon}\\
& \ln\dfrac{s'}{s'-s}\,=\,\ln\dfrac{1}{1-z}\,=\,-\int d\zeta\,\dfrac{\theta(\zeta-1)}{z-\zeta+i\epsilon},
\end{align}
where $\theta(x)$ is the step function. Notice $\,\mathrm{Im}\left[\int_{m^2}^{+\infty}ds'\ln(m^2/s')\,r_1(s')\right]\,=\,0 $. Then through the spectral representation, $\kappa(\zeta)$ defines another function by $f(s/s')=\int d\zeta\,\kappa(\zeta)/(s/s'-\zeta+i\varepsilon)$. Therefore $\overline{R}_1(s)$, the spectral function of $q_1(p^2)$, is given by, 
\begin{align}
\overline{R}_1(s)&=\,-\dfrac{1}{\pi}\mathrm{Im}\{q_1(s+i\epsilon) \}\nonumber\\[1mm]
& =\, \dfrac{3\alpha}{4\pi}\dfrac{m^2}{s}\theta(s-m^2)-\dfrac{\alpha\xi}{4\pi}\theta(s-m^2)\nonumber\\[1mm]
& \quad-\dfrac{3\alpha}{4\pi}\left[\dfrac{4}{3}sr_1(s)-\int_{m^2}^{s}ds'\dfrac{s'}{s}r_1(s) \right]\nonumber\\[1mm]
&\quad -\dfrac{\alpha\xi}{4\pi}\int_{m^2}^{s}ds'r_1(s').\label{eq:def_R1_bar}
\end{align}

Similarly with the help of
\begin{align}
\dfrac{1}{s}\left(-1+\dfrac{s'}{s}\ln\dfrac{s'}{s'-s}\right)\,&=\,\dfrac{1}{s'z}\left(-1+\dfrac{1}{z}\ln\dfrac{1}{1-z}\right)\,\nonumber\\[1mm]
&=\,-\dfrac{1}{s'}\int \dfrac{d\zeta}{\zeta^2}\,\dfrac{\theta(\zeta-1)}{z-\zeta+i\epsilon}\, ,
\end{align}
the spectral function of $q_2(p^2)$ is given by
\begin{align}
\overline{R}_2(s)& =-\dfrac{1}{\pi}\mathrm{Im}\{q_2(s+i\epsilon) \}\nonumber\\[1mm]
& =\dfrac{3\alpha}{4\pi}\dfrac{m}{s}\theta(s-m^2)-\dfrac{\alpha\xi}{4\pi}\dfrac{m^3}{s^2}\theta(s-m^2)\nonumber\\[1mm]
& \quad  -\dfrac{3\alpha}{4\pi}\left[\dfrac{4}{3}r_2(s)-\int_{m^2}^{s}ds'\dfrac{1}{s}r_2(s') \right]\nonumber\\[1mm]
&\quad -\dfrac{\alpha\xi}{4\pi}\int_{m^2}^{s}ds'\dfrac{s'}{s^2}r_2(s').\label{eq:def_R2_bar}
\end{align}

Before deriving the SDE for functions $r_{j}(s)$, we need to calculate the imaginary part of inhomogeneous terms in Eq.~\eqref{eq:SDE_r1_OnShell}. Although the value of $\theta(s-m^2)$ at $s=m^2$ remains undetermined, the linear combination ${\,\overline{R}_1(m^2)-m\overline{R}_2(m^2)}$ is free of such ambiguity:
\begin{align*}
\overline{R}_1(m^2)\,& =\,\lim\limits_{s\rightarrow m^2}\,\left[\dfrac{3\alpha}{4\pi}\theta(s-m^2)-\dfrac{\alpha\xi}{4\pi}\theta(s-m^2) \right]\nonumber\\[1.5mm]
& \quad -\dfrac{\alpha}{\pi}m^2r_1(m^2),
\end{align*}
\begin{align}
\overline{R}_2(m^2)\,&=\,\lim\limits_{s\rightarrow m^2}\left[\dfrac{3\alpha}{4\pi}\dfrac{1}{m}\theta(s-m^2)-\dfrac{\alpha\xi}{4\pi}\dfrac{1}{m}\theta(s-m^2) \right]\nonumber\\[1.5mm]
&\quad -\dfrac{\alpha}{\pi}r_2(m^2),\nonumber\\[1.5mm]
\overline{R}_1(m^2)&-m\overline{R}_2(m^2)\,=\,-\dfrac{\alpha}{\pi}[m^2r_1(m^2)-mr_2(m^2)].
\end{align}
Therefore, taking the imaginary parts of Eq.~\eqref{eq:SDE_rj_OnShell} produces
\begin{subequations}
\begin{align}
sr_1(s)+\overline{R}_1(s)& \,=\,\left(1-\dfrac{\alpha}{\pi}\right)mr_2(s)\nonumber\\
&\quad \,+2\left(1-\dfrac{\alpha}{\pi}\right)[m^2r_1(m^2)-mr_2(m^2)],\\
r_2(s)+\overline{R}_2(s)& =\left(1-\dfrac{\alpha}{\pi}\right)mr_1(s).
\end{align}
\end{subequations}
At $s\,=\,m^2$, we have
\begin{subequations}
\begin{align}
& \quad[m^2r_1(m^2)-mr_2(m^2)]\left(1-\dfrac{\alpha}{\pi}\right)\nonumber\\
& \quad +\dfrac{(3-\xi)\alpha}{4\pi}\lim\limits_{s\rightarrow m^2}\theta(s-m^2)\,\nonumber\\
&=\,2[m^2r_1(m^2)-mr_2(m^2)]\left(1-\dfrac{\alpha}{\pi}\right),\\[2mm]
&\quad  [r_2(m^2)-mr_1(m^2)]\left(1-\dfrac{\alpha}{\pi}\right)\nonumber\\
&+\dfrac{(3-\xi)\alpha}{4\pi}\dfrac{1}{m}\lim\limits_{s\rightarrow m^2}\theta(s-m^2)\,=\,0,
\end{align}
\end{subequations}
or equivalently, 
\begin{align}
&\quad \left(1-\dfrac{\alpha}{\pi}\right)[m^2r_1(m^2)-mr_2(m^2)]\nonumber\\
& =\,\dfrac{(3-\xi)\alpha}{4\pi}\lim\limits_{s\rightarrow m^2}\theta(s-m^2),\label{eq:SDE_ini_cond_1}
\end{align}
with $m\neq 0$.
Adopting $\,{\lim\limits_{s\rightarrow m^2}\theta(s-m^2)\,=\,1}\,$ produces non-trivial solutions. The choice of this particular limit of the $\theta$-functions  will be explained later in this section. 
The equations for functions $r_{j}(s)$ then become
\begin{subequations}\label{eq:SDE_LG_OnShell}
\begin{align}
\;\, sr_1(s)+\overline{R}_1(s)\,&=\,\left(1-\dfrac{\alpha}{\pi}\right)mr_2(s)+\dfrac{(3-\xi)\alpha}{2\pi} \\
\quad r_2(s)+\overline{R}_2(s)\,&=\,\left(1-\dfrac{\alpha}{\pi}\right)mr_1(s),
\end{align}
\end{subequations}
where $\overline{R}_j(s)$ are specified by Eqs.~(\ref{eq:def_R1_bar},~\ref{eq:def_R2_bar}).
\section{The Frobenius method for Eq.~\eqref{eq:ODE_fj}\label{ss:Frobenius}}
The homogeneous part of Eq.~\eqref{eq:ODE_f1} can be solved using the Frobenius method. Let us start with
\begin{align}
f_1(x)\,&=\,x^{a+1}\sum_{n=0}^{+\infty}c_nx^n\,\nonumber\\
&=\,c_0\,x^{a+1}~_2F_1(a+1,a+1;2a+2;-x).\label{eq:f_1_homogeneous}
\end{align} 
This inspires the following decomposition
\begin{equation}
f_1(x)\,=\,x^{a+1}\,\Phi_1(x),
\end{equation}
where $x\,=\,s/m^2-1$. Eq.~\eqref{eq:ODE_f1} then becomes
\begin{align}
x(x+1)\dfrac{d^2}{dx^2}\Phi(x)+&\left[2a+2+(2a+3)x\right]\,\dfrac{d}{dx}\,\Phi(x)\nonumber\\
& +(a+1)^2\,\Phi(x)\,=\,2a(a+1).\label{eq:ODE_phi_1}
\end{align}
The homogeneous part of Eq.~\eqref{eq:ODE_phi_1} is the hypergeometric differential equation for $\Phi$ with variable $-x$. After finding a particular solution to the inhomogeneous equation, the general solution to Eq.~\eqref{eq:ODE_phi_1} with finite initial conditions is
\begin{equation}
\Phi(x)\,=\,c_0\,_2F_1(a+1,a+1;2a+2;-x)+\dfrac{2a}{a+1}.
\end{equation}
Therefore, the solution for Eq.~\eqref{eq:ODE_f1} is
\begin{equation}
f_1(x)\,=\,x^{a+1}\left[c_0\,_2F_1(a+1,a+1;2a+2,-x)+\dfrac{2a}{a+1} \right].\label{eq:f1_exact}
\end{equation}
Substituting Eq.~\eqref{eq:f1_exact} into Eq.~\eqref{eq:df1_f2} gives the solution to Eq.~\eqref{eq:ODE_f2} directly:
\begin{equation}
f_2(x)\,=\,-c_0\,\dfrac{(a+1)}{a}x^a~_2F_1(a,a+1;2a+2;-x).\label{eq:f2_exact}
\end{equation}
Based on Eqs.~(\ref{eq:f1_exact},~\ref{eq:f2_exact}) for functions $f_{j}(x)$, the following solutions are obtained using Eq.~\eqref{eq:rj_decomposition}: 
\begin{subequations}\label{eq:rj_exact}
\begin{align}
&\quad r_1(s)\,\nonumber\\
&=\,\dfrac{c_0}{s}\,_2F_1\left(a+1,a+1;2a+2;-\dfrac{s-m^2}{m^2}\right)+\dfrac{2a}{(a+1)s},\label{eq:r1_exact}\\[1.5mm]
&\quad r_2(s)\,\nonumber\\
&=\,-c_0\,\dfrac{(1+a)}{am}~_2F_1\left(a+1,a+2;2a+2;-\dfrac{s-m^2}{m^2}\right).\label{eq:r2_exact}
\end{align}
\end{subequations}
The parameter $c_0$ can be determined by the the initial condition Eq.~\eqref{eq:SDE_ini_cond_1}. 

Given that ${_2F_1(a+1,a+1;2a+2,0)\,=\,1}$, and ${~_2F_1(a+1,a+2;2a+2;0)\,=\,1}$, we have
\begin{equation}
c_0+\dfrac{2a}{a+1}\lim\limits_{s\rightarrow m^2}\theta(s-m^2)+c_0\,\dfrac{a+1}{a}\,=\,2a\lim\limits_{s\rightarrow m^2}\theta(s-m^2),
\end{equation}
where the implicit dependence of the inhomogeneous term on the limit of the $\theta$-function has been restored. Nontrivial solutions are obtained with ${\lim\limits_{s\rightarrow m^2}\theta(s-m^2)\,=\,1}$. Then 
\begin{equation}
c_0\,=\,\dfrac{2a^3}{(2a+1)(a+1)}\label{eq:c0_nontrivial}
\end{equation}
gives the solution Eq.~\eqref{eq:r12_2F1}.
If, instead, the boundary value $\lim\limits_{s\rightarrow m^2}\theta(s-m^2)\,=\,0$ had been chosen, only homogeneous parts of Eq.~\eqref{eq:rj_exact} survive. As a result, the only viable choice of $c_0$ is $0$, which produces trivial solutions.

Because Eq.~\eqref{eq:ODE_f1} is a second order ordinary differential equation, there is another linearly independent solution of Eq.~\eqref{eq:f1_exact}. The corresponding $r_1(s)$ is
\begin{align}
\quad r_1(s)\,&=\,\dfrac{2a}{(a+1)s}+\dfrac{c}{s}\,
\left(\dfrac{s}{m^2}-1\right)^{-2a-1}\nonumber\\
&\hspace{1.4cm}\times\,_2F_1(-a,-a;-2a;1-s/m^2).\label{eq:r1_alternative}
\end{align}
An interesting observation is that apart from different definitions of coupling parameter $a$, the homogeneous solution in Eq.~\eqref{eq:r1_alternative} is identical to the corresponding $r_1(s)$ from Eq.~\eqref{eq:rW_Delbourgo} of Ref.~\cite{Delbourgo:1977jc} based on the renormalization scheme given by Eq.~\eqref{eq:Z2_Zm_Delbourgo}. However, Eq.~\eqref{eq:r1_alternative} is not taken within the on-shell renormalization scheme because it does not satisfy the finite boundary condition Eq.~\eqref{eq:SDE_ini_cond_1}. Specifically, it corresponds to a propagator function having a different $p^2\rightarrow m^2$ behavior from the on-shell propagators. 
\bibliographystyle{unsrt}
\bibliography{manuscript_2F1_bib}
\end{document}